\documentclass[12pt]{article}

\usepackage{scicite}

\usepackage{times}

\usepackage{amsmath}
\usepackage{amsfonts}
\usepackage{amssymb}
\usepackage{graphicx}
\usepackage{ulem}

\usepackage{bm,color,amsmath,amssymb,latexsym,graphicx,psfrag,accents,float}
\usepackage{amscd,dsfont,wasysym,mathrsfs,mathtools}
\usepackage{multirow}
\usepackage{dcolumn}
\usepackage{xcolor}
\usepackage{comment}

\usepackage{booktabs} 

\usepackage{lineno}

        \definecolor{AAcolor}{rgb}{0.7,0.1,0.4}

\title{Reply to: Low-frequency quantum oscillations in LaRhIn$_5$: Dirac point or nodal line?}

\author
{Chunyu Guo,${}^{1,2\ast\dagger}$ A. Alexandradinata,${}^{3,4,5\ast\dagger}$ Carsten Putzke,${}^{1,2}$ Amelia Estry,${}^{2}$ Teng \\ Tu,${}^{6}$ Nitesh Kumar,${}^{7}$ Feng-Ren Fan,${}^{7}$ Shengnan Zhang,${}^{8,9}$ Quansheng Wu,${}^{8,9}$ \\Oleg V. Yazyev,${}^{8,9}$ Kent R. Shirer,${}^{7}$ Maja D. Bachmann,${}^{7,10}$ Hailin Peng,${}^{6}$ Eric\\ D. Bauer,${}^{11}$ Filip Ronning,${}^{11}$ Yan Sun,${}^{7}$ Chandra Shekhar,${}^{7}$  Claudia Felser,${}^{7}$ \\ Philip J. W. Moll,${}^{1,2\dagger}$\\
\normalsize{${}^{1}$Max Planck Institute for the Structure and Dynamics of Matter, Hamburg, Germany}\\ 
  \normalsize{${}^{2}$Laboratory of Quantum Materials (QMAT), Institute of Materials (IMX),}\\ 
  \normalsize{\'{E}cole Polytechnique F\'{e}d\'{e}rale de Lausanne (EPFL), CH-1015 Lausanne, Switzerland}\\
\normalsize{${}^{3}$ Department of Physics and Santa Cruz Materials Center, University of California, }\\ 
\normalsize{Santa Cruz, California, USA}\\
 \normalsize{${}^{4}$Institute for Condensed Matter Theory, University of Illinois at Urbana-Champaign,}\\ \normalsize{ Urbana, Illinois 61801, USA}\\
  \normalsize{${}^{5}$Department of Physics, University of Illinois at Urbana-Champaign, }\\ \normalsize{Urbana, Illinois 61801, USA}\\
  \normalsize{${}^{6}$Center for Nanochemistry, Beijing National Laboratory for Molecular Sciences (BNLMS),}\\ \normalsize{College of Chemistry and Molecular Engineering, Peking
University, Beijing 100871, China.}\\
 \normalsize{${}^{7}$Max Planck Institute for Chemical Physics of Solids, 01187 Dresden, Germany}\\
 \normalsize{${}^{8}$Chair of Computational Condensed Matter Physics (C3MP), Institute of Physics (IPHYS),}\\ \normalsize{\'{E}cole Polytechnique F\'{e}d\'{e}rale de Lausanne (EPFL), CH-1015 Lausanne, Switzerland}\\
 \normalsize{${}^{9}$National Centre for Computational Design and Discovery of Novel Materials MARVEL, }\\ \normalsize{\'{E}cole Polytechnique F\'{e}d\'{e}rale de Lausanne (EPFL), CH-1015 Lausanne, Switzerland}\\
 \normalsize{${}^{10}$School of Physics and Astronomy, University of St Andrews, St Andrews KY16 9SS, UK}\\
 \normalsize{${}^{11}$Los Alamos National Laboratory, Los Alamos, New Mexico 87545, USA}\\
 \\
 \normalsize{$^\ast$These authors contributed equally to this work.}\\
 \normalsize{$^\dagger$Corresponding authors: chunyu.guo@epfl.ch(C.G); aalexan7@illinois.edu(A.A.);}
 \\
 \normalsize{philip.moll@epfl.ch(P.J.W.M.)}\\}

\date{}

\begin{document}

\baselineskip24pt

\maketitle

\section*{}
We thank G.P. Mikitik and Yu.V. Sharlai for contributing this note\cite{MattArisingMikitik} and the cordial exchange about it. First and foremost, we note that the aim of our paper\cite{OurNC} is to report a methodology to diagnose topological (semi)metals using magnetic quantum oscillations. Thus far, such diagnosis has been based on the phase offset of quantum oscillations, which is extracted from a “Landau fan plot”. A thorough analysis of the Onsager--Lifshitz--Roth quantization rules has shown that the famous $\pi$-phase shift can equally well arise from orbital- or spin magnetic moments in topologically trivial systems with strong spin-orbit coupling or small effective masses\cite{phiB_Aris_PRX}. Therefore, the “Landau fan plot” does not by itself constitute a proof of a topologically nontrivial Fermi surface. In the paper at hand\cite{OurNC}, we report an improved analysis method that exploits the strong energy-dependence of the effective mass in linearly dispersing bands. This leads to a characteristic temperature dependence of the oscillation frequency which is a strong indicator of nontrivial topology, even for multi-band metals with complex Fermi surfaces. Three materials, Cd$_3$As$_2$, Bi$_2$O$_2$Se and LaRhIn$_5$ served as test cases for this method. Linear band dispersions were detected for Cd$_3$As$_2$, as well as the $F$ $\approx$ 7 T pocket in LaRhIn$_5$. \\
We reiterate that the temperature dependence of the oscillation frequency only encodes the band dispersion at the Fermi level, but is not indicative of the band structure away from the Fermi level. A linear band dispersion can arise from either a 3D Dirac point or a nodal-line degeneracy; both scenarios lead to the same $T^2$-coefficient of the oscillation frequency, if such a coefficient is properly normalized. Mikitik and Sharlai have previously proposed the nodal-line scenario for the small pocket of LaRhIn$_5$, which they show to be compatible with our temperature-dependent frequency. In our paper, we raised the possibility of the Dirac-point scenario because our ab-initio calculations suggested that the spin-orbit coupling is not as weak as was assumed in \cite{La115_Miktik_PRL}. However, we did not commit to either the Dirac-point or the nodal-line scenario because our ab-initio calculations did not exactly match the experimental findings. Independent of which scenario holds, we reiterate that our proposed methodology of identifying linear band dispersions is not in question, and this linearity has been robustly identified in LaRhIn$_5$; the only controversy lies in  the  nature of the band crossing away from the Fermi level.\\
Such a controversy deserves some attention. Significant activities in the field of heavy fermions are devoted to our understanding of Ce(Co,Rh,Ir)In$_5$, a fantastically rich sandbox to test ideas about quantum criticality, unconventional superconductivity and exotic correlated electron states\cite{Ce115_JDT_review}. It is commonly accepted that LaRhIn$_5$, lacking the critical 4f$^1$ electron of Ce, is a trivial metal which is only studied to subtract trivial phonon contributions from data on their Ce-bearing siblings. Mikitik and Sharlai spotlight\cite{MattArisingMikitik} an essential problem: even without the complexity of 4f physics, forming an accurate description of the electronic structure of such complex multi-band metals remains a challenge for ab-initio calculations. Yet if we do not understand the uncorrelated baseline well, little hope remains for formulating a good quantitative understanding of the electronic structure of CeRhIn$_5$.\\

In an early and seminal work, Mikitik and Sharlai\cite{La115_Miktik_PRL} identified a Berry-phase contribution to the phase offset of de Haas-van Alphen (dHvA) oscillations observed by Goodrich et al\cite{La115_goodrich_PRL}. As replotted in the Fig. 2 of \cite{MattArisingMikitik}, the Goodrich data plausibly match Mikitik and Sharlai's model prediction (blue curve), which is based on a nodal line enclosed by `neck-like' Fermi surface with a minimal cross-sectional area. For comparison, Mikitik-Sharlai also plotted (in red) the model prediction based on a 3D Dirac point enclosed by an isotropic Fermi surface. Most strikingly, the blue curve fits well to the ultra-quantum limit of Goodrich's data, while the red curve does not.\\

While Goodrich's measurement of dHvA oscillations was with a fixed field orientation, we have measured the angle-dependent Shubnikov--de Haas (SdH) oscillations of the same Fermi pocket, by tilting the field in the a-c crystallographic plane. We found that the oscillation frequency remained between 6 to 9~T for the field orientations that we measured; moreover, the effective mass varies only by 20\% over the 90$ ^{\circ}$ range between the a and c axes. Indeed, no evidence was found for the highly anisotropic Fermi pocket that typically encircles a nodal line. The weak angle-dependence of the SdH oscillation frequency supports the model of a 3D Dirac point enclosed by an isotropic Fermi surface.\\ 

Unfortunately, our SdH data does not fit all the predictions based on an isotropic Fermi surface. In particular, the left-right asymmetry of the SdH oscillation peaks is visible by eye and should reflect the left-right asymmetry of the density-of-states for  magnetic energy levels. This is formalized by a phase offset of $+\pi/4$ (or $-\pi/4$) for a minimal (or maximal) cross-sectional area, as is well-known from the Lifshitz--Kosevich formula. Our SdH data can only be fitted with a $+\pi/4$ offset, which is consistent with a minimal-cross-section Fermi surface that encloses either a nodal line or a 3D Dirac point. Such a minimal cross section invariably results in an anisotropic Fermi surface, which is at odds with the weak angle-dependence of the SdH frequency.\\

To recapitulate, three plausible models have been presented by us and Mikitik-Sharlai: (i) a nodal line with a minimal cross-section, (ii) a Dirac point with a maximal cross-section, and (iii) a Dirac point with a minimal cross-section. None of the three models are obviously compatible with both the Goodrich dHvA and our SdH data. A further discomforting piece of this puzzle is the failure of DFT-based methods, which do not predict any nodal lines or Dirac points in the vicinity of the Fermi level.\\ 


Let us speculate on possible resolutions of this puzzle. It is possible a finer angular resolution for the SdH frequency would identify singularities that were missed in the current data; it would also be useful to vary the field orientation over a different crystallographic plane. Furthermore, close inspection uncovers some concerns about the original data by Goodrich et al. In the low-field region, their data shows a pronounced diamagnetic signal which is highly unusual for a non-magnetic intermetallic which is commonly dominated by Pauli paramagnetism. One may hypothesize about a potential artefact arising from the pulsed-field gradient magnetometry used in this study. Given the high conductivity of LaRhIn$_5$, dynamic diamagnetism from eddy currents may contribute or even dominate the signal, which then may not be interpreted as magnetization. A dominant SdH character would also explain the identical left-right asymmetry compared to our SdH measurements. This calls for a revisit of the magnetic properties and dHvA oscillations in LaRhIn$_5$, ideally in a static magnetic field.

We hope this exchange inspires further work to clarify the electronic structure of LaRhIn$_5$. There is a clear need to refine the electronic modeling and for new experiments to approach this problem. Something unusual seems to be hidden in what most consider the trivial complexity of LaRhIn$_5$.

\bibliographystyle{Nature}

\clearpage

\noindent \textbf{Data Availability.} 
Data that support the findings of this study is deposited to Zenodo with the access link: https://doi.org/10.5281/zenodo.7724832.\\\\

\noindent \textbf{Code Availability.} 
Mathematica code used for the Lifshitz-Kosevich fit of temperature-dependent quantum oscillations can be
found via the access link:  https://doi.org/10.5281/zenodo.5482689.\\

\noindent \textbf{Author Contributions.} \\
High quality LaRhIn$_5$ Crystals were synthesized and characterized by  F.R., E.D.B.. The magnetotransport measurements were performed by C.G., C.P., A.E., K.R.S., M.B., P.J.W.M.. A.A. developed, applied and described the theoretical framework, and the analysis of experimental results has been done by C.G. and C.P.. All authors were involved in writing the paper. 

\noindent \textbf{Competing Interests.} The authors declare no competing interests.\\

\section*{Acknowledgements} 
\noindent \textbf{Funding.} A.A. was supported initially by the Yale Postdoctoral Prize Fellowship, and subsequently by the Gordon and Betty Moore Foundation EPiQS Initiative through Grant No. GBMF 4305 and GBMF 8691 at the University of Illinois. S.Z, Q.W. and O. V. Y. acknowledge support by NCCR Marvel and the Swiss National Supercomputing Centre (CSCS) under Projects No. s832 and No. s1008. M.D.B. acknowledges studentship funding from EPSRC under grant no EP/L015110/1. E.D.B. and F.R. were supported by the U.S. DOE, Basic Energy Sciences, Division of Materials Sciences and Engineering. C.P. and P.J.W.M. acknowledge funding by the European Research Council (ERC) under the European Union's Horizon 2020 research and innovation programme ("MiTopMat" - grant agreement No. 715730). This project was funded by the Swiss National Science Foundation (Grants  No. PP00P2\_176789).\\


\clearpage

\begin{figure}[t]
	\includegraphics[width=0.95\columnwidth]{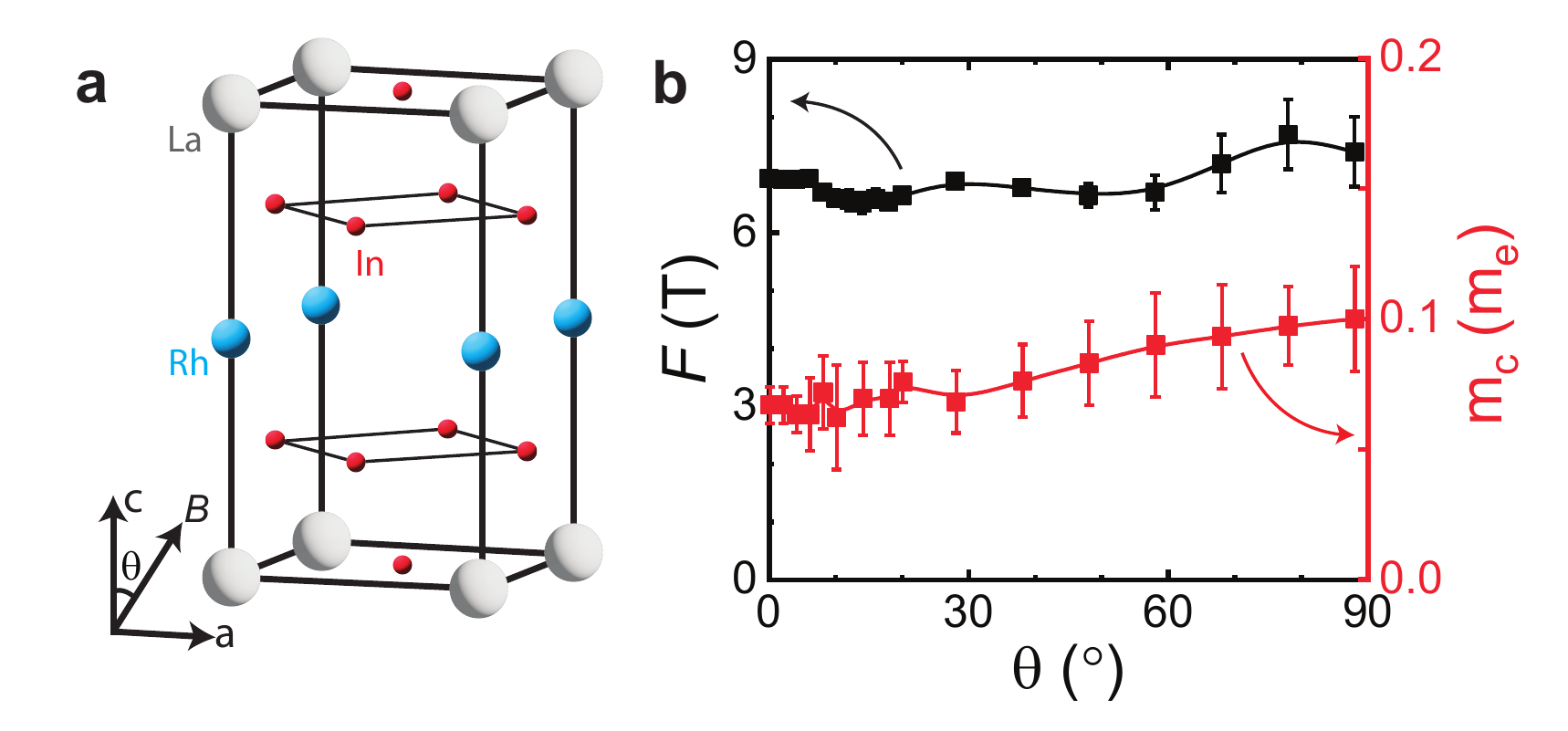}
	\caption{\textbf{Angular dependence of quantum oscillation frequency and cyclotron mass in LaRhIn$_5$.} (a) Illustration of Crystal structure and field orientation, the field angle is defined as the angle between the crystalline c--direction and magnetic field. (b) Shubnikov--de Haas oscillation frequency spectrum as a function of field angle, as well as the cyclotron mass. Both parameters are found to be highly angle insensitive, suggesting a nearly isotropic and closed Fermi pocket. All error bars are determined by the standard error generated by the non-linear regression Lifshitz--Kosevich fitting procedure as described in \cite{OurNC}.}
\end{figure}

\end{document}